# Shadow Behavior of the Quantum-Corrected Schwarzschild Black Hole Immersed in Holographic Quintessence

Sara Saghafi[1], Kourosh Nozari[2]
[1]Department of Theoretical Physics, University of Mazandaran, Babolsar.
Email: s.saghafi@umz.ac.ir
[2]Department of Theoretical Physics, University of Mazandaran, Babolsar.
Email: knozari@umz.ac.ir

**Abstract.** In this paper, we aim to explore the impact of the Planck scale corrections and the Holographic quintessence on the shadow behavior of non-rotating black holes. To do this, we consider the quantum-corrected Schwarzschild black hole surrounded by the quintessence field inspired by the Kazakov-Solodukhin and the Kiselev ideas, and we call this combination the Kazakov-Solodukhin-Kiselev (KSK) black hole. We conclude that the quintessence field as the candidate of dark energy in the black hole can be interpreted as Holographic quintessence. To find the geodesic equations of the black hole, we employ the Hamilton-Jacobi approach and also, the Carter procedure. We discover that the size of the shadow of this black hole, which depends on its central mass, is also determined by the Planck scale effects and Holographic quintessence.

## 1 Introduction

Recently, a vast number of astrophysical data, such as observations of type Ia supernovae [1] and so on, show us that currently, the Universe is experiencing an accelerated phase in its expansion, which is widely believed that it is due to some kind of negative-pressure form of energy, known as dark energy. The simplest candidate for dark energy within the structure of General Theory of Relativity (GR) proposed by Einstein is the cosmological constant [2], which is related to the vacuum energy with a constant energy density and pressure, and a parameter of equation of state $\omega_\Lambda = -1$. An alternative proposal for dark energy is the dynamical scenario to describe the nature of dark energy. This dynamical proposal is characterized by some scalar field mechanism, which suggests that the negative-pressure form of energy is provided by a scalar field. One of the most simple, famous models of dynamical dark energy is the quintessence scalar field [3] with the parameter of equation of state $\omega_q > -1$ for the spatially homogeneous case. Due to the assumption of its homogeneity, the field is considered to be extremely light.

Another alternative to describe the nature of the dark energy, is arisen from a quantum gravity outcome, known as the holographic principle [4], firstly proposed by 't Hooft [5] in the black hole physics. According to the holographic principle, the entropy of a system scales with its surface area, not its volume. Based on quantum field theory [6], a short-distance cut-off is related to a long-distance (IR) cut-off, because of the black hole formation limit. If the quantum vacuum energy is due to a short-distance cut-off, then the total energy in a region of size $L$ has not to exceed a black hole mass of the same size, i.e., $(L^3 \rho_h \leq L M_p^2)$. Therefore, by taking the whole universe into account, the vacuum energy associated with the holographic principle can be considered as dark energy, so-called holographic dark energy [7,8]. The holographic dark energy density for the largest $L$ is $\rho_h = 3\lambda^2 M_p^2 L^{-2}$ in which $\lambda^2$ is a constant, $M_p^{-2} = 8\pi G$ is the Planck mass, and $G$ is the Newtonian gravitational constant. It is shown that for $\lambda \geq 1$, the holographic dark energy can be explained by quintessence field, known as holographic quintessence, with a parameter of equation of state in range $-1 < \omega_{hq} < -1/3$ [7,8].

Recently, Kiselev [9] considered the quintessence field in the background of Schwarzschild black hole using quintessence stress–energy tensor with the additivity and linearity conditions to derive a Schwarzschild-like solution of GR surrounded by quintessence dark energy. So, in the range $-1 < \omega_{hq} < -1/3$, one can take the Kiselev's solution into account as a black hole solution in the background of holographic quintessence. On the other hand, by considering the quantum effects at the Planck scale, Kazakov and Solodukhin [10] in 1994 modified Schwarzschild black hole, so that they removed its point-like singularity. The Kazakov-Solodukhin's black hole has a central 2-sphere of radius $a$ rather a central point-like singularity due to the presence of quantum effects, and $a$ is the quantum parameter of the setup, which is of the order of Planck's length $l_p$. It is possible to combine the Kiselev and Kazakov-Solodukhin solutions [11] to gain a regular Schwarzschild back hole in the background of the holographic quintessence with $\omega_{hq} = -2/3$ as a special case in the range $-1 < \omega_{hq} < -1/3$. We named black hole as the Kazakov-Solodukhin-Kiselev (KSK) black hole.

In this paper, we aim to study the shadow behavior of the KSK black hole to find how quantum effects of the spacetime and also, the holographic dark energy affects the shadow of black holes. We know that if a black hole is in front of a luminous background, it will produce a shadow, which is a ring of light around a region of darkness. Such a ring of light is created by matter circling at the very edge of the event horizon. For a non-rotating black hole, the shape of the shadow, which is circular, together with its size are determined by the black hole's mass. The rest of the paper is organized as follows. In Section 2 we introduce the line element of the KSK black hole, briefly. In Section 3 we study the motion of photons in the KSK spacetime, and then we investigate the shadow behavior for this. In Section 4 we have a discussion and review our results. Finally, in Section 5 we end with a brief conclusion. In the whole of this paper, we set $G = c = \hbar = 1$.

## 2 Quantum-Corrected Schwarzschild Black Hole in the Background of Holographic Quintessence

Combining the Kiselev [9] and Kazakov-Solodukhin [10] ideas as the procedure performed in Ref. [11], one can find the following line element

$$ds^2 = -f(r)dt^2 + \frac{dr^2}{f(r)} + r^2(d\theta^2 + \sin^2\theta \, d\phi^2) \,, \tag{1}$$

where

$$f(r) = -\frac{2M}{r} + \frac{1}{r}\sqrt{r^2 - a^2} - \frac{\sigma}{r^{3\omega_{hq}+1}} \,, \tag{2}$$

in which $M$ is source mass and $\sigma$ is a positive normalization constant corresponding with the holographic quintessence. Due to the presence of quantum effects, the line element (1) depicts a regular Schwarzschild black hole surrounded by holographic quintessence for which a central 2-sphere of radius $a$ substitutes for the central point-like singularity of the Schwarzschild black hole. Also, the regular black hole experiences the late time accelerated expansion of the Universe because of the presence of holographic quintessence in the background as a candidate of dark energy. Now, we just need to put $\omega_{hq} = -2/3$ in Eq. (2) to obtain the metric coefficient of the KSK black hole as follows

$$\tilde{f}(r) = -\frac{2M}{r} + \frac{1}{r}\sqrt{r^2 - a^2} - \sigma r \,. \tag{3}$$

Considering the metric coefficient (3), in the following, we want to study the shadow behavior of the KSK black hole.

## 3 Shadow Behavior of the KSK black hole

The Lagrangian of a test particle with mass $m$ in the spacetime background of the KSK black hole is as follows

$$\mathcal{L} = \frac{1}{2} g_{\mu\nu} \dot{x}^\mu \dot{x}^\nu = \frac{1}{2}\left[-\tilde{f}(r)\dot{t}^2 + \frac{1}{\tilde{f}(r)}\dot{r}^2 + r^2\dot{\theta}^2 + r^2 \sin^2\theta\, \dot{\phi}^2\right], \quad (4)$$

in which "dot" denotes derivation with respect to an affine parameter, $\tau$ and $g_{\mu\nu}$ is the metric tensor of KSK black hole. The canonically conjugate momentum's components can be found out as

$$P_t = \tilde{f}(r)\dot{t} = E, \qquad P_r = \frac{1}{\tilde{f}(r)}\dot{r}, \qquad P_\theta = r^2\dot{\theta}, \qquad P_\phi = r^2 \sin^2\theta\, \dot{\phi} = L, \quad (5)$$

in which $E$ and $L$ as the energy and the angular momentum of the test particle, respectively are two constants of the motion arising from two Killing vectors, $\partial_t$ and $\partial_\phi$ of the KSK black hole.

To investigate the motion and orbits of photon, we make use of the Hamilton-Jacobi approach and also, we consider the Carter method [12] to formulate the geodesic equations for the KSK black hole. The Hamilton-Jacobi equation is to the form

$$\frac{\partial S}{\partial \tau} = -\frac{1}{2} g^{\mu\nu} \frac{\partial S}{\partial x^\mu} \frac{\partial S}{\partial x^\nu}, \quad (6)$$

where $S$ is the Jacobi action. We Assume a separable solution for Jacobi action as

$$S = \frac{1}{2} m^2 \tau - Et + L\phi + S_r(r) + S_\theta(\theta). \quad (7)$$

For photon, we have $m = 0$. Inserting Eq. (7) into the Hamilton-Jacobi equation (6) results in

$$0 = \frac{E^2}{\tilde{f}(r)} - \tilde{f}(r)\left(\frac{\partial S_r}{\partial r}\right)^2 - \frac{1}{r^2}\left(\frac{L^2}{\sin^2\theta} + \mathcal{K} - L^2 \cot^2\theta\right) - \frac{1}{r^2}\left(\left(\frac{\partial S_\theta}{\partial \theta}\right)^2 - \mathcal{K} + L^2 \cot^2\theta\right), \quad (8)$$

in which $\mathcal{K} = \left(r^2\dot{\theta}\right)^2 + \frac{L^2}{\sin^2\theta}$ is the Carter constant. Therefore, one can recast Eq. (8) as the following two separated equations

$$r^4 \tilde{f}^2(r)\left(\frac{\partial S_r}{\partial r}\right)^2 = r^4 E^2 - r^2 \tilde{f}(r)(\mathcal{K} + L^2), \quad (9)$$

$$\left(\frac{\partial S_\theta}{\partial \theta}\right)^2 = \mathcal{K} - L^2 \cot^2\theta. \quad (10)$$

From Eqs. (5), (9), and (10), one can find the complete null geodesic equations for the KSK black hole as follow

$$\dot{t} = \frac{E}{\tilde{f}(r)}, \qquad \dot{\phi} = \frac{L}{\sin^2\theta}, \quad (11)$$

$$r^2 \dot{r} = \pm\sqrt{\mathcal{R}} = \pm\sqrt{[r^4 E^2 - r^2 \tilde{f}(r)(\mathcal{K} + L^2)]} \quad (12)$$

$$r^2 \dot{\theta} = \pm\sqrt{\Theta} = \pm\sqrt{[\mathcal{K} - L^2 \cot^2\theta]} \quad (13)$$

where plus (minus) is for outgoing (ingoing) radial direction of photon's motion.

One can define two impact parameters $\xi = L/E$ and $\eta = \mathcal{K}/E^2$ to analyze the properties of photon's motion around the KSK black hole. On the other hand, it is well known that the boundaries of the shadow of a black hole is determined by the unstable null circular orbits. To find this, one can rewrite the radial null geodesic equation for the KSK black hole as

$$\left(\frac{dr}{d\tau}\right)^2 + V_{eff}(r) = 0, \qquad V_{eff}(r) = \frac{1}{r^2}\tilde{f}(r)(\mathcal{K} + L^2) - E^2, \quad (14)$$

in which $V_{eff}(r)$ is the effective potential for radial photon's motion. The unstable null circular orbits are available when the effective potential becomes maximum, which occurs in the following conditions

$$V_{eff} = \frac{dV_{eff}}{dr}\bigg|_{r=r_o} = 0, \qquad \mathcal{R} = \frac{d\mathcal{R}}{dr}\bigg|_{r=r_o} = 0, \qquad (15)$$

where $r_o$ known as photon sphere radius is the certain value of $r$ for which $V_{eff}$ becomes maximum. From conditions introduced in Eq. (15), one can find that $r_o$ is the solution of the following equation

$$r_o \tilde{f}'(r_o) - 2\tilde{f}(r_o) = 0. \qquad (16)$$

One can recast the effective potential, $V_{eff}(r)$ and the function $\mathcal{R}(r)$ in terms of two impact parameters, $\eta$ and $\xi$ as follow

$$V_{eff}(r) = E^2 \left[ \frac{1}{r^2} \tilde{f}(r)(\eta + \xi^2) - 1 \right], \qquad \mathcal{R}(r) = E^2 \left[ r^4 - r^2 \tilde{f}(r)(\eta + \xi^2) \right]. \qquad (17)$$

Therefore, inserting Eq. (17) into Eq. (15), one can find

$$\eta + \xi^2 = \frac{4r_o^2}{2\tilde{f}(r_o) + r_o \tilde{f}'(r_o)}. \qquad (18)$$

We compute the values $\eta + \xi^2$ and $r_o$ for some different values of $\sigma$, $\omega_{hq}$, and $a$ in Table 1 to investigate the variation of $\eta + \xi^2$ in terms of $r_o$. In Table 1, the case of $a = 0 = \sigma$ is for Schwarzschild black hole, just for comparison. We should note that in the considered unit setup, i.e., $G = c = \hbar = 1$, the photon sphere radius, $r_o$ has the dimension of length, while the quantity $\eta + \xi^2$ has the dimension of length square. From Table 1, one can see that increasing $\sigma$ leads to increase $r_o$ and the quantity $\eta + \xi^2$ for a fixed $a$. Also, for a fixed $\sigma$, increasing $a$, results in increasing $r_o$ and the quantity $\eta + \xi^2$.

Table 1. Values of $r_o$ and $\eta + \xi^2$ for different values of $\sigma$, $\omega_{hq}$, and $a$.

|  | $\sigma = 0$ | | $\sigma = 0.05$ | | $\sigma = 0.1$ | |
| --- | --- | --- | --- | --- | --- | --- |
|  | $r_o$ | $\eta + \xi^2$ | $r_o$ | $\eta + \xi^2$ | $r_o$ | $\eta + \xi^2$ |
| $a = 0$ | 3 | 9 | 3.266 | 14.552 | 3.675 | 41.609 |
| $a = 1$ | 3.312 | 9.473 | 3.611 | 15.941 | 4.084 | 57.110 |
| $a = 2$ | 3.592 | 9.906 | 4.472 | 19.999 | 5.152 | 282.78 |

To characterize the real shadow seen on the observer's frame (sky), one should use the celestial coordinates, $\alpha$ and $\beta$ [13]. These coordinates make it easier to study the shape of the black hole shadow. The celestial coordinates can be defined as follow

$$\alpha = \lim_{r_o \to \infty} \left( -r_o^2 \sin \theta_o \frac{d\phi}{dr} \right), \qquad \beta = \lim_{r_o \to \infty} \left( r_o^2 \frac{d\theta}{dr} \right), \qquad (19)$$

where $\theta_o$ is the inclination angle between the black hole's z-axis and the sight line from source to observer. To be precise, the celestial coordinates are two apparent perpendicular distances of the shadow as seen from the axis of symmetry, and its projection on the equatorial plane, respectively. Utilizing the deduced null geodesic equations, one can assume the observer on the equatorial plane ($\theta = \pi/2$) to read the celestial coordinates as

$$\alpha = -\xi, \qquad \beta = \pm\sqrt{\eta}. \qquad (20)$$

Using Eq. (20), one can read Eq. (18) as

$$\alpha^2 + \beta^2 = \eta + \xi^2 = \frac{4r_o^2}{2\tilde{f}(r_o) + r_o\tilde{f}'(r_o)} = R_s^2, \tag{21}$$

in which $R_s$ is the perfect circle's radius of the shadow since the KSK black hole is a non-rotating one. This radius approximately characterizes the shadow size. Fig. 1 shows the illustration of the shadow of the KSK black hole in the celestial plane $(\alpha, \beta)$ for some different values of $a$ and $\sigma$. The dashed red and dashed orange circles are respectively for $\sigma = 0.05$ and $\sigma = 0.1$ with $a = 1$. Also, blue dot-dashed and purple dot-dashed circles are respectively for $\sigma = 0.05$ and $\sigma = 0.1$ with $a = 2$. From Fig. 1 we again see that for larger values of $a$, the radius of the shadow increases. This is because of strengthening the quantum effects, which grows the central 2-sphere, results in increasing the size of the black hole and its shadow. On the other hand, Fig. 1 depicts that increasing $\sigma$, leads to increasing the size of the black hole shadow. Also, the black dotted curve is for Schwarzschild black hole, just for comparison.

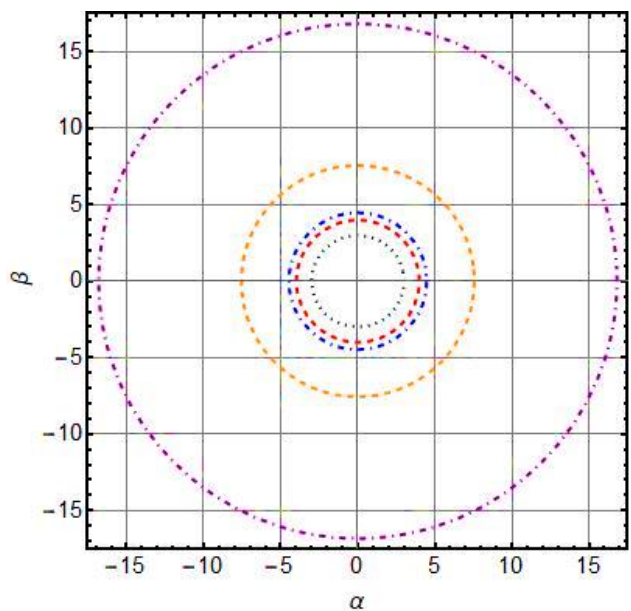

Figure 1. Shadow of KSK black hole in the celestial plane for different values of $a$ and $\sigma$.

## 4 Results and Discussion

We studied the shadow behavior for the regular Schwarzschild black hole surrounded by holographic quintessence called the KSK black hole inspired by the Kiselev and Kazakov-Solodukhin's ideas. We aimed to investigate how quantum effects and holographic quintessence will affect the shadow of a black hole. Especially, since the KSK black hole is a regular one due to the presence of quantum effects, studying its shadow and then comparing the outcomes with the observational data will help us to know if quantum effects play something special in background of spacetime. We found that increasing the quantum effects through increasing the quantum parameter leads to increase the shadow radius. This situation is the same for increasing the effect of holographic quintessence.

## 5 Conclusions

The main conclusion of the paper is that quantum effects play an important role in the background spacetime, so that they directly change the shadow behavior of a black hole. Also, the presence of holographic quintessence as a candidate for dark energy, change the shadow of a black hole, too. So, we can say that, the shadow size of a black hole is determined by background quantum effects and dark energy ingredient of the Universe, in addition to the mass of the black hole.